# 从不确定性原理到确定性判则
## —不确定性原理非统计诠释的证伪与统计诠释的突破

段德龙

（中国科学院空天信息创新研究院，北京 100190）

**摘要**

不确定性原理的物理诠释分歧一直存在。为消除该分歧对量子信息技术发展的制约与困扰，本文从回溯海森堡不确定性原理数学关系式的原始推导、物理涵义出发，分析了爱因斯坦光子箱思想实验，研究了与统计诠释对立的不确定性原理正统诠释，考察了广义作用图景和引力作用图景下的不确定性关系式。通过分析量子力学量的统计分布，获得了电磁作用图景下非统计诠释不确定性关系式被破坏的结果；通过研究光子箱思想实验，发现了玻尔当年论证的逻辑矛盾；通过考察作用图景集合，提出了微观粒子确定性力学状态的描述方法；根据能量守恒原理对氢原子跃迁辐射过程的分析，得到了引力作用图景下的不确定性关系式的下限远小于现有电磁作用图景下该关系式下限的结果。本文的研究揭示了非统计诠释所指称的量子个体几率性的来源和非统计诠释自身所蕴含的内在逻辑矛盾，明确了统计诠释的适用范围，解决了以不确定性原理为标志的量子力学物理诠释分歧问题；同时，本文的研究揭示了微观粒子力学状态具有的客观确定性，发现了确定其客观力学状态的极限判则。本研究有助于增进对微观世界量子物理现象的理解，对于考察量子信息技术的物理基础，亦可提供相应的参考。

**关键词**：量子力学，不确定性原理，哥本哈根诠释，光子箱，引力波，量子信息

**PACS**：03.65.Ca, 03.65.Ta, 01.70.+w, 01.65.+g, 95.85.Sz, 03.67.-a

# From the uncertainty principle to the deterministic rule
## --The falsification of the non-statistical interpretation of the uncertainty principle and the breakthrough of the statistical interpretation

Duan De-Long

(Aerospace Information Research Institute, Chinese Academy of Sciences; Beijing, 100190)

**Abstract:** In order to eliminate the divergence of the uncertainty principle in the physical interpretation of quantum mechanics, this paper traces the original derivation and the physical meaning of the mathematical relationship of Heisenberg uncertainty principle, analyzes the thought experiment of Einstein's photon box, and studies the gravitational scene and the set of action scenarios. By analyzing the statistical distribution of quantum mechanical quantities, the result of the destruction to uncertainty relation under the non-statistical interpretation in the electromagnetic interaction scenarios is obtained; through analyzing of the photon box thought experiment, the logical contradiction of Bohr's argument was discovered; by examining the set of interaction scenarios, a description method for determining the mechanical state of microscopic particles was put forward; according to the analysis of the hydrogen atom transition radiation process, basing on the principle of conservation of energy, the lower limit of the uncertainty relation in the gravitational scene is much smaller than that in the existing electromagnetic scene, which mean the lower limit of uncertainty principle is broken. The research in this article has received an affirmative answer to Einstein's God does not play dice with the Universe, and may be helpful to enhance the understanding of quantum physical phenomena in the microscopic world, even renders some theoretical support to investigating the physical basis of quantum technology.

# 1 引言

作为现代物理学的两大支柱之一,量子力学在应用方面获得了科学史上前所未有的巨大成功,它不仅深刻改变了物理学,也彻底影响和改变了人类社会[1][2][3][4]。与此同时,发端于不确定性原理的量子力学物理诠释纷争长期以来一直存在[2][5],并且伴随量子信息技术的快速布局,以不确定性原理为焦点的量子力学物理诠释分歧问题对量子技术发展的制约与困扰日愈明显。

不确定性关系,亦即不确定性原理,玻尔、泡利等称之为量子力学理论的基石。它的提出,被认为是科学史上的一个重大成就。作为量子力学与经典力学之间本质差异的标志,不确定性原理展现出量子力学的几率性和非决定性(indeterminacy)与经典力学的决定论(determinism)之间的根本不同,并直接引发了爱因斯坦与玻尔间的世纪论战。从与量子力学逻辑自洽性等价的不确定性原理逻辑一贯性,到EPR理论所针对的量子力学完备性,二人持续论战了几十年。直到今天,量子力学非统计诠释关于微观世界的几率性描写图景与爱因斯坦坚持的经典因果决定性精确化描述之间的分歧依然如旧[1][2][3][4][5][6][7][8]。

对于不确定性关系式的数学推导,物理学界并无异议,根本分歧在于不确定性原理的物理诠释,其物理诠释分为以下两个流派[1]:

**Ⅰ 非统计诠释**:不确定性关系是对个体量子系统的描述,量子系统的正则共轭变量不能同时被精确确定,其不确定度之积下限为$\hbar/2$[6]。这是量子力学主流诠释,亦被称为量子力学正统诠释、哥本哈根诠释,其核心是强调量子力学的个体几率性,放弃了自伽利略时代起物理学一直坚守的因果决定性。

**Ⅱ 统计诠释**——爱因斯坦、薛定谔所坚持的诠释:不确定性关系描述的是全同制备的量子系统所构成的系综,其正则共轭变量统计分布的标准差之积下限为$\hbar/2$。其实质是坚持而不是放弃经典的因果决定性精确化描述图景。

玻尔、爱因斯坦索尔维论战之后,海森堡、薛定谔、德布罗意、马根瑙、布洛欣采夫和张汉良、关洪、黄湘友以及小泽正直等,都在该方面进行过有益的探索,但仍未能消除其物理诠释上的根本分歧。为此,下文将①回顾不确定性关系式的原始推导与物理涵义,对非统计诠释下的该关系式进行概率分析,研究其数学物理上的逻辑自洽性;②回顾玻尔关于光子箱思想实验的论证,探究其论证的逻辑一贯性;③分析不确定性关系式与波函数数学形式随广义作用图景基本作用量'常量'趋向零极限时的演化,考察量子几率性的来源并分析精确描述微观粒子力学状态的可能;④推导引力作用图景下的不确定性原理数学关系式,研究现行量子力学与不确定性关系式的适用范围。

# 2 不确定性原理的提出与非统计诠释

量子力学数学形式体系的构建超前于其物理诠释,为了解决其数学形式体系所面临的两个问题:①量子理论是否容许一个粒子的位置与速度在一给定的时刻只能以有限的精度被确定,②理论允许的精度与实验测量中获得的最佳精度是否相容,海森堡于1927年3月发表了《论量子理论的运动学和力学的直观内容》[1][6],提出不确定性原理并通过量子力学数学形式体系导出了关系式(1):

$$\Delta q \Delta p \geq \hbar/2 \tag{1}$$

式中$\Delta q \equiv \sqrt{\overline{(\Delta q)^2}} = \sqrt{\overline{(x-\bar{x})^2}}$、$\Delta p \equiv \sqrt{\overline{(\Delta p)^2}} = \sqrt{\overline{(p_x-\overline{p_x})^2}}$，并在[6]中通过γ射线显微镜及电子的单狭缝位置测定等思想实验，对不确定性关系式进行了物理诠释[1][2]。定义不确定性关系式是量子理论对同时测量共轭力学量精确度的限制，是非统计诠释的核心。非统计诠释强调"任何一个确切的观测结果（精度）都要遵守不确定度关系"[8]，坚持精度限制的物理来源是测量操作导致的无法避免的干扰[6][9][10][11][12][13]，认为不确定性原理保护着量子力学的逻辑自洽性[10]。

非统计诠释是目前绝大部分中外量子力学/理论力学教科书以及自然哲学专著和量子理论科普著作使用的标准诠释[1][2][6][14][15][16]，是量子通信与量子计算的理论基础（后文以关系式$\Delta q \Delta p \geq \hbar/2$ (2)表示非统计诠释下的不确定性原理关系式）。

# 3 不确定性原理的分析探讨

为了解决不确定性原理在物理诠释上的根本分歧问题，研究其适用范围，本章将首先对微观粒子量子力学量的统计分布与非统计诠释下的不确定关系式进行符合性分析，考察该关系式的逻辑一贯性（3.1节）；接着分析玻尔对光子箱思想实验的论证，研究玻尔论证的逻辑自洽性（3.2节）；尔后构造作用图景集合，考察对基本作用量'常量'取极限时微观粒子力学状态的描述（3.3.1/2节），最后对不确定性关系式的引力作用图景表达进行分析讨论（3.3.3节）。

## 3.1 不确定性关系式非统计诠释的数学分析

本节将分析量子力学量的统计分布与非统计诠释下的不确定性关系式的符合性，以考察不确定性关系式非统计诠释的逻辑自洽性。

### 3.1.1 力学量的统计分布与非统计诠释的符合性分析

为了实现量子力学量的统计分布与非统计诠释的不确定性关系式的符合性分析，与[6]的定义与假设相同，记函数$\psi(q')$、$\Psi(p')$分别为电子的力学量$q$、$p$分布的概率幅，设$q$为高斯分布、记其概率幅为$\psi(q')$，则由傅里叶变换可得$p$亦为高斯分布、其概率幅记为$\Psi(p')$，$\psi(q')$、$\Psi(p')$表达式如下

$$\psi(q') = \frac{1}{(2\pi)^{1/4}\sqrt{(\Delta q)}} \cdot e^{-\frac{(q'-\bar{q})^2}{4(\Delta q)^2} \frac{2\pi i}{h} pq'}$$

$$\Psi(p') = \int \frac{1}{\sqrt{h}} e^{\frac{i}{\hbar}q'p'} \psi(q') dq' \quad (3)$$

① 定义$\Delta q' = q' - \bar{q}$为单次测量$q$对$\bar{q}$的偏差，表示对$q$的单次测量精度。记$q'$满足$q' \in (\bar{q} - \Delta q, \bar{q} + \Delta q)$，即$\Delta q' < \Delta q$的$(q', p')$概率为$\eta_{q'}$：

$$\eta_{q'} = \{\int_{\bar{q}-\Delta q}^{\bar{q}+\Delta q} |\psi(q')|^2 dq'\} / \{\int |\psi(q')|^2 dq'\} \quad (4)$$

其中$\Delta q$为(2)式中$q$为高斯分布时所定义的$\Delta q$。

由定积分性质可知，若在[a,b]上有函数$f(x) \geq 0$成立，则

$$\int_a^b f(x) dx \geq 0$$

必定成立。所以，由$q' \in (\bar{q} - \Delta q, \bar{q} + \Delta q)$时$|\psi(q')|^2 > 0$以及$\int |\psi(q')|^2 dq' = 1$可得：

$$\eta_{q'} = \{\int_{\bar{q}-\Delta q}^{\bar{q}+\Delta q} |\psi(q')|^2 dq'\} / \{\int |\psi(q')|^2 dq'\} > 0 \quad (5)$$

② 因$q$和$p$为高斯分布，所以(2)式中$\Delta q$、$\Delta p$满足$\Delta q \Delta p = \hbar/2$，式中$\Delta p$为(2)式$p$为高斯分布时所定义的$\Delta p$。记满足$\{p' \in (\bar{p}-\Delta p, \bar{p}+\Delta p), \Delta q \Delta p = \hbar/2\}$的$(q', p')$的概率为$\eta_{p'}$， $(q', p')$ 满足 $\{q' \in (\bar{q}-\Delta q, \bar{q}+\Delta q), p' \in (\bar{p}-\Delta p, \bar{p}+\Delta p), \Delta q \Delta p = \hbar/2\}$ 的概率为 $\eta_{q'p'}$，则：

$$\eta_{p'} = \int_{\bar{p}-\Delta p}^{\bar{p}+\Delta p} \Psi(p')\Psi^*(p')dp'$$

$$= \int_{\bar{p}-\Delta p}^{\bar{p}+\Delta p} \left(\int \frac{1}{\sqrt{h}} e^{\frac{i}{\hbar}q'p'} \psi(q')dq'\right) \left(\int \frac{1}{\sqrt{h}} e^{\frac{i}{\hbar}q'p'} \psi(q')dq'\right)^* dp' > 0 \qquad (6)$$

$$\eta_{q'p'} = \int_{\bar{p}-\Delta p}^{\bar{p}+\Delta p} \left(\int_{\bar{q}-\Delta q}^{\bar{q}+\Delta q} \frac{1}{\sqrt{h}} e^{\frac{i}{\hbar}q'p'} \psi(q')dq'\right) \left(\int_{\bar{q}-\Delta q}^{\bar{q}+\Delta q} \frac{1}{\sqrt{h}} e^{\frac{i}{\hbar}q'p'} \psi(q')dq'\right)^* dp'$$

$$= \int_{\bar{p}-\Delta p}^{\bar{p}+\Delta p} \left|\int_{\bar{q}-\Delta q}^{\bar{q}+\Delta q} \frac{1}{\sqrt{h}} e^{\frac{i}{\hbar}q'p'} \psi(q')dq'\right|^2 dp' \qquad (7)$$

由(7)式与定积分性质可得

$$\eta_{q'p'} = \int_{\bar{p}-\Delta p}^{\bar{p}+\Delta p} \left|\int_{\bar{q}-\Delta q}^{\bar{q}+\Delta q} \frac{1}{\sqrt{h}} e^{\frac{i}{\hbar}q'p'} \psi(q')dq'\right|^2 dp' > 0 \qquad (8)$$

$$\eta_{q'p'} > 0 \qquad (9)$$

③ 由(7)、(8)式可得，当满足条件 $\{q' \in (\bar{q}-\Delta q, \bar{q}+\Delta q)、p' \in (\bar{p}-\Delta p, \bar{p}+\Delta p)、\Delta q \Delta p = h/4\pi\}$，亦即下式

$$\Delta q' \Delta p' < \hbar/2 \qquad (10)$$

存在时，(9)式成立

$$\eta_{q'p'} > 0 \qquad (9)$$

概率分析结果(9)&(10)式显然与非统计诠释的预言相矛盾，非统计诠释关于"任何一个确切的观测结果（精度）都要遵守的不确定度关系式"[9](2)失效、不成立。

### 3.1.2 对不确定性原理非统计诠释失效的探讨

3.1.1 条的概率分析结果之所以与非统计诠释的预言相矛盾，原因在于：

[6]推导不确定性关系式时，海森堡不仅用到了坐标、动量两种表象的统计性波函数，还用到了表象变换和对变换函数的统计诠释；而且对γ射线显微镜进行的定性物理分析，同样用到了隐含的统计性关系——爱因斯坦-德布罗意关系。若没有统计性波函数与德布罗意关系，就无法得到不确定性关系式(1)。作为量子力学数学形式体系的一个直接推论-即通过狄拉克-约尔当变换理论获得的一个结论[1][2][3][6]，不确定性关系式(1)表示共轭力学量标准差之间的约束关系，是统计诠释的等价数学表述。

玻尔认为单纯数学形式本身的明晰性对于一个物理命题没什么价值,完备的物理解释应绝对高于其数学形式体系[1][14]，海森堡在导出不确定性关系式(1)之后，将式中的均方差 $\sqrt{(\Delta q)^2}$、$\sqrt{(\Delta p)^2}$ 定义为电子位置和动量的不确定度用以表示测量精确度，把该数学关系式诠释为同时测量两个不同物理量的精确度[6]之间要遵从的约束关系(2)。

不确定性关系式(1)中的均方差 $\sqrt{(\Delta q)^2}$、$\sqrt{(\Delta p)^2}$ 表征共轭力学量 $q$、$p$ 值的弥散程度、呈现的是 $q$、$p$ 值的统计分布特征[1][2][3]，二者都必需通过对大量高精度的测量结果进行统计分析才能获得，无法仅通过单次测量就得到[5]；而海森堡在关系式(2)中却明确用其表示测量精确度——精确度为 $q$、$p$ 获得值对真（期望）值的偏离度，属于单次测量的个体特性量。海森堡将关系式(1)中的 $\Delta q$、$\Delta p$ 重新诠释定义为测量精确度，使其失去了在统计理论中标准差的固有数学涵义，在概念上造成了 $\Delta q$、$\Delta p$ 的数学/物理涵义发生由均方根差→不确定度→测量精确度的改变，内涵前后不一。无容置疑，这违背逻辑一贯性原则，因此 3.1.1 条进行概率分析时得到了关系式(2)遭破坏、非统计诠释不成立的结果。

同时，非统计诠释迄今为止还没有获得任何真实物理实验的直接验证支持[1][5]。不仅如此，曾为其提供诠释支撑的海森堡γ-射线显微镜和爱因斯坦光子箱等几个思想实验[1][6]，经

再分析（参见本文 3.2 条及[17]），同样得到了与 3.1.1 数学证明相一致的——不确定性关系式(2)遭破坏的研究结果。

作为量子力学统计诠释的等价数学表述，不确定性关系式对与标准差来源不同的单次测量的精度没有约束、不存在任何限制，也不能对力学量 $p$、$q$ 的单次测量精确度之积$\Delta q'\Delta p'$作出任何限定，单次测量精确度没有上下限。统计诠释并不否定微观粒子可同时具有确定的位置与动量；而且，统计意义上的不确定性关系并非量子力学所独有，经典理论同样不乏其对应存在。[3][17][18][19][20]

所以，非统计诠释缺乏逻辑一贯性，不确定性关系与现行量子力学理论仅在统计诠释下才能对微观粒子的力学状态作出恰当的描述。

### 3.2 对光子箱思想实验的再分析

玻尔在第六届索尔维会议上对光子箱思想实验的反论证，驳倒了爱因斯坦关于非统计诠释不能逻辑自洽的观点，奠定了非统计诠释在量子力学中的主流、正统地位[1][15]。本节将对此进行回顾分析，考察玻尔论证的逻辑自洽性。

### 3.2.1 玻尔与爱因斯坦关于光子箱辩论的回顾

在 1930 年 10 月的布鲁塞尔索尔维会议上，爱因斯坦为驳倒玻尔坚持的不确定性原理非统计诠释，提出了光子箱思想实验，认为光子发射时间 T 和光子能量 E 的测量彼此独立，互不干扰，二者的测量精确度没有相互制约，所以该思想实验将破坏时间-能量不确定性关系式$\Delta T \cdot \Delta E \geq \hbar$，从而证明不确定性原理与量子力学的非统计诠释不能逻辑自洽[1][8][15]。玻尔当即对此进行了反论证[15]：

在能量为 E 的光子逸出时，光子箱由此获得一个向上的动（冲）量 $p$

$$p \leq T\frac{E}{c^2}g \tag{11}$$

式(11)中 T 为光子箱自光子释放开始至达到新的平衡位置所需时间。光子箱获得的动量 $p$ 的不确定度$\Delta p$为

$$\Delta p \leq T\frac{\Delta E}{c^2}g \tag{12}$$

光子逸出前后，光子箱的两次平衡位置之差$\Delta x$与$\Delta p$有关，$\Delta x$、$\Delta p$满足(13)式

$$\Delta p \geq \frac{\hbar}{\Delta x} \tag{13}$$

由(12)、(13)可得

$$\hbar \leq T\frac{\Delta E}{c^2}\Delta x g \tag{14}$$

T 的不确定度$\Delta T$由位置不确定度$\Delta x$(引力势变化导致钟表快慢变化)决定，根据引力红移公式

$$\frac{\Delta T}{T} = g\frac{\Delta x}{c^2} \tag{15}$$

将(15)代入(14)即得出时间-能量不确定性关系式

$$\Delta E \cdot \Delta T \geq \hbar \tag{16}$$

玻尔的论证，爱因斯坦无法反驳，也为当年的与会者所接受[1][8][15]，并被认为是"对量子力学复杂性的一个特别严明的分析"。此后，爱因斯坦未再公开质疑不确定性原理与量子力学的逻辑自洽性，但实际上他始终未认同非统计诠释。并且，玻尔自 1930 年索尔维会议结束到 1962 年去世，曾反复回到对光子箱论证的思考上[1][8]。

### 3.2.2 对玻尔论证的再分析

玻尔当年的论证，被认为是"非常清晰地说明了正确诠释量子力学所必须依靠的物理基础"。那么它是否在逻辑上确实无懈可击、真正能够证明非统计诠释的逻辑自洽性？下面从玻尔的论证逻辑和对玻尔论证结论的逻辑反推两个方面进行分析。

3.2.2.1 玻尔的论证，存在以下几个问题

① 由(11)式到(12)式的推导存在错误

由动（冲）量定义

$$p = T\frac{E}{c^2}g \tag{17}$$

忽略高阶小量、可得下式：

$$\Delta p = \Delta T\frac{E}{c^2}g + T\frac{\Delta E}{c^2}g \tag{18}$$

玻尔从动量 $p$ 表达式(11)得出动量不确定度$\Delta p$ 表达式(12)的论证不成立。

② 玻尔论证过程中数学符号涵义前后不一致的问题

在"光子…满足(13)式"中的$\Delta x$ 表示光子箱两次平衡位置之差；在"T 的不确定…(15)式"中的$\Delta x$ 却用来表示位置不确定度。事实上，光子箱两次平衡位置之差与其位置不确定度，二者并不相同。

③ 玻尔论证存在的逻辑循环问题

为了证明时间-能量的不确定性关系，玻尔在"光子…满足(13)式"中先验地断定位置-动量不确定性关系式成立，并引入其变式$\Delta p \geq \frac{h}{\Delta x}$，这在论证逻辑上属不证自引，是循环论证。

④ 关于"光子…满足(13)式"是否成立的问题

设光子能量为 E，则由质能方程可得光子箱释放光子前后两次平衡位置之差$\Delta x$为

$$\Delta x = \frac{E}{kc^2}g \tag{19}$$

位置差$\Delta x$取决于光子能量 E 与光子箱牵引簧的弹性系数 k，与来源不明的$\Delta p$并无关系，$\Delta x$、$\Delta p$ 二者不受(13)式的约束，所以玻尔论证依据"光子…满足(13)式"不成立。

⑤ 光子箱称重时两次平衡位置间过渡时间测量的不确定度问题

记引力红移时钟变慢效应导致光子箱称重时间$T_b$（$T_b$即为 3.2.1 节(11)~(16)式中玻尔论证使用的时间 T；另记光子箱释放光子开闭快门所需的时间为 $T_a$）的变化量为$\Delta T_b$，由引力红移公式可知：

$$\Delta T_b = T_b\frac{g\Delta x}{c^2} \tag{20}$$

$\Delta T_b$是可通过上式准确计算的时间变化量，不可作为时间不确定度。

记光子箱的位置不确定度为$\Delta x'$，$\Delta x'$由称重时测量光子箱平衡位置的极限精度决定。显然，一个具有物理意义的位置测量必有$\Delta x' < \Delta x$成立。位置不确定度$\Delta x'$将引起引力势变化的不确定，进而导致钟表快慢变化的不确定，由此带来时间 $T_b$ 测量的不确定度为

$$\Delta T_b' = T_b\frac{g\Delta x'}{c^2} \tag{21}$$

因$\Delta x' < \Delta x$，所以时间$T_b$测量的不确定度$\Delta T_b' < \Delta T_b$。因此，玻尔论证依据的时间不确定度不正确。

3.2.2.2 依照玻尔的论证逻辑对玻尔论证的反论证

① 设能量 E 与时间 $T_b$为高斯分布，且假定玻尔的论证成立，可得下式

$$\Delta E \cdot \Delta T_b = \hbar \tag{22}$$

② 记$\Delta x_1$为光子箱释放光子时间$T_a$内的位移，记光子箱开启快门时刻 $t_1$、光子箱关闭快门时刻$t_2$、光子箱到达新的平衡位置并最终完成称重的时刻为 $t_3$，$t_1<t_2<t_3$，显然$\Delta x_1(v(t)>0$，$\Delta x_1=\int_{t_1}^{t_2} v(t)dt)$小于称重过程光子箱的总位移 $(\Delta x=\int_{t_1}^{t_2} v(t)dt + \int_{t_2}^{t_3} v(t)dt)$，即 $\Delta x_1 < \Delta x$。

由公式(15)，可得光子箱释放光子时间 $T_a$（$T_a=t_2-t_1$）因引力红移所导致的时间变化量$\Delta T_a$为

$$\Delta T_a = g\frac{\Delta x_1}{c^2}T_a \tag{23}$$

因光子箱的位置不确定度而带来的时间 $T_a$ 测量的不确定度为$\Delta T_a'$：

$$\Delta T_a' = g\frac{\Delta x'}{c^2}T_a \tag{24}$$

由于光子箱释放光子时间$T_a$极短，光子箱称重过程的时间$T_b$($T_b=t_3-t_1$)可任意长[1][8]，即$T_a < T_b\{\Delta x_1 < \Delta x\}$，所以$\Delta T_a' < \Delta T_b'\{\Delta T_a < \Delta T_b\}$得下式：

$$\Delta E \cdot \Delta T_a' < \Delta E \cdot \Delta T_b' < \Delta E \cdot \Delta T_b \qquad (25)$$

$$\{\Delta E \cdot \Delta T_a < \Delta E \cdot \Delta T_b \qquad (26)\}$$

因①已假设 E 满足高斯分布，关系式$\Delta E \cdot \Delta T_b = \hbar$成立，所以根据(22)式可以预言光子箱释放的光子到达远方的时间-能量不确定性关系式满足下式

$$\Delta E \cdot \Delta T_a' < \hbar \qquad (27)$$

$$\{\Delta E \cdot \Delta T_a < \hbar \qquad (28)\}$$

(27){(28)}式的结果破坏时间-能量不确定性关系式(16)与(22)，玻尔的索尔维论证被证伪（注：本条{...}为依[15]的玻尔论证逻辑的推导）。

玻尔在 1930 年索尔维会议上关于光子箱思想实验的论证，曾被认为是成功证明不确定性原理与量子力学非统计诠释逻辑自洽的范例[1][15]。但是上文的回顾分析表明，玻尔当年的论证显然不成立，不确定性原理与量子力学的非统计诠释不能逻辑自洽（注：对文献[1][8]的分析研究，亦得到了同样的结论）。

### 3.3 非电磁作用图景的讨论

3.1&3.2 条证明了非统计诠释在逻辑上不能自洽，统计诠释是不确定性原理与量子力学的合理物理诠释。那么，统计诠释下不确定性原理与现行量子力学能否普适地对（各种作用场景下的）微观粒子力学状态进行描述？为此，本节将首先在广义作用图景下推导共轭力学量的标准差约束关系式；尔后通过构造作用图景集合，考察其基本作用量'常量'取极限时微观粒子运动状态的描述；最后根据能量守恒原理，通过引入引力作用量'常量'，在对氢原子电磁辐射过程的分析中实现对不确定性关系式引力作用图景表达的考察。

### 3.3.1 广义作用图景下共轭力学量标准差约束关系式的推导

为实现广义作用图景下对微观粒子力学状态的描述，依据玻恩对易关系量子化思想和基本力学量海森堡方程，参照[6][21][22]，设待定广义作用图景为 $X(\varepsilon_x)$，定义其基本作用量'常量'为$\varepsilon_x$。在 $X(\varepsilon_x)$ 作用图景下，任一粒子处于某有限空间，记该空间的三个坐标轴分别为 x, y, z。为简化分析，仅考虑一维情况、只讨论三个空间坐标中的 x 坐标，令粒子沿 x 方向运动，$p_x$ 为该粒子的动量。设坐标波函数为$\psi(x)$，动量波函数为$\Psi(p_x)$，以标准偏差表示坐标和动量的不确定度

$$\Delta x = \sqrt{\overline{(x-\bar{x})^2}}, \quad \Delta p_x = \sqrt{\overline{(p_x - \overline{p_x})^2}}$$

假定 x 和 $p_x$ 的平均值皆为零，考虑下列明显成立的不等式

$$\int_{-\infty}^{\infty} \left| \alpha x \psi + \frac{d\psi}{dx} \right|^2 dx \geq 0 \qquad (29)$$

其中$\alpha$是为任意实常数。

因为：

$$\int x^2 |\psi|^2 dx = (\Delta x)^2 \qquad (30)$$

$$\int \left( x \frac{d\psi^*}{dx} \psi + x \psi^* \frac{d\psi}{dx} \right) dx = \int x \frac{d|\psi|^2}{dx} dx = -\int |\psi|^2 dx = -1 \qquad (31)$$

$$\int \frac{d\psi^*}{dx} \frac{d\psi}{dx} dx = -\int \psi^* \frac{d^2\psi}{dx^2} dx = \frac{1}{\varepsilon_x^2} \int \psi^* \hat{p}_x^2 \psi dx = \frac{1}{\varepsilon_x^2} (\Delta p_x)^2 \qquad (32)$$

所以

$$\alpha^2 (\Delta x)^2 - \alpha + \frac{1}{\varepsilon_x^2} (\Delta p_x)^2 \geq 0 \qquad (33)$$

关于$\alpha$的抛物型二次三项式(33)对所有的$\alpha$都大于等于零，因而其判别式必须为非正的实数，所以

$$\Delta x \Delta p_x \geq \frac{1}{2} \varepsilon_x \qquad (34)$$

$\Delta x \Delta p_x$乘积的最小可能值为$\frac{1}{2}\varepsilon_x$，此时波函数呈下列形式：

$$\psi(x) = \frac{1}{(2\pi)^{1/4}\sqrt{(\Delta x)}} exp\left(\frac{i}{\varepsilon_x}p_x x - \frac{x^2}{4(\Delta x)^2}\right) \tag{35}$$

其中$p_0$和$\Delta x$是常数，$\Delta x$为标准偏差，得高斯分布的概率函数

$$|\psi|^2 = \frac{1}{\sqrt{(2\pi)}\cdot\Delta x} exp\left(-\frac{(x-\bar{x})^2}{2(\Delta x)^2}\right) \tag{36}$$

动量表象中的波函数为

$$\Psi(p_x) = \frac{1}{\sqrt{2\pi\varepsilon_x}}\int_{-\infty}^{\infty}\psi(x)e^{(-i/\varepsilon_x)xp_x}dx \tag{37}$$

其概率为

$$|\Psi|^2 = \frac{1}{2\pi\varepsilon_x} \times \exp\left(-\frac{(p_x - \overline{p_x})^2}{4(\Delta p_x)^2}\right) \tag{38}$$

动量概率分布$|\Psi|^2$同样为高斯分布，其标准偏差为$\Delta p_x = \varepsilon_x/2\Delta x$，即$x$、$p_x$的标准差约束关系式为：

$$\Delta x \Delta p_x = \frac{1}{2}\varepsilon_x \tag{39}$$

### 3.3.2 微观粒子力学状态的确定描述

为了考察能否获得微观粒子力学状态的确定性描述，构造广义作用图景集合$\{X_i\}$（i=1,2,3…n…），$\{X_i\}$对应的基本作用量'常量'集合$\{\varepsilon_{xi}\}$的元素序列有下式的关系：

$$\varepsilon_{x1} > \varepsilon_{x2} > \varepsilon_{x3} > ... > \varepsilon_{xn} > ... \to 0 \tag{40}$$

即：

$$\lim_{n\to\infty}\varepsilon_{xn} = 0 \tag{41}$$

设作用图景$X_i$下微观粒子位置$x_i$为高斯分布，则动量$p_{xi}$亦为高斯分布且$x_i$、$p_{xi}$标准差之积满足

$$\Delta x_i \Delta p_{xi} = \frac{1}{2}\varepsilon_{xi} \tag{42}$$

根据式(40)与(42)，下式成立

$$\Delta x_1 \Delta p_{x1} > \Delta x_2 \Delta p_{x2} > \Delta x_3 \Delta p_{x3} > ...\Delta x_n \Delta p_{xn} > ... \to 0 \tag{43}$$

即：

$$\lim_{n\to\infty}\Delta x_n \Delta p_{xn} = 0 \tag{44}$$

由作用量定义可得：

$$\lim_{n\to\infty}\Delta x_n = 0 \tag{45}$$

$$\lim_{n\to\infty}\Delta p_{xn} = 0 \tag{46}$$

当$n \to \infty$时，微观粒子的动量与位置波函数(24)与(25)式演化为两个δ函数[28][29]：

$$\lim_{\Delta x_n \to 0}|\psi| = \lim_{\Delta x_n \to 0}\left|\frac{1}{(2\pi)^{1/4}\sqrt{(\Delta x_n)}}exp\left(-\frac{i}{\varepsilon_{xn}}\overline{p_{xn}}x_n - \frac{x_n^2}{4(\Delta x_n)^2}\right)\right| = \delta(x = \bar{x}, \Delta x_n = 0) \tag{47}$$

$$\lim_{\varepsilon_{xn} \to 0}|\Psi| = \lim_{\varepsilon_{xn} \to 0}\left|\frac{1}{(2\pi)^{1/4}\sqrt{\varepsilon_{xn}}}\int_{-\infty}^{\infty}\psi(x_n)exp\left(-\frac{i}{\varepsilon_{xn}}x_n p_{xn}\right)dx_n\right| = \delta(p_x = \overline{p_x}, \varepsilon_{xn} = 0) \tag{48}$$

通过广义作用图景描述分析中蜕化为δ函数的(47)、(48)两式，即得到微观粒子确定的位置与动量值，进而可准确获知微观粒子的力学状态。由此极限判则可断定，微观粒子具有客观、确定、并且可被准确描述的力学状态。

### 3.3.3 引力作用图景下的不确定性关系

统计诠释为不确定性原理与量子力学的合理物理诠释，那么它能否恰当地描述引力作用图景下微观粒子的力学状态？本小节根据能量守恒原理，通过引入引力作用量'常量'，对氢原子发光过程进行分析，实现不确定性关系式的引力作用图景表达，考察海森堡不确定性

关系式(1)与现有量子力学描述的适用性。

自然界存在的四种基本作用力，都由粒子传递：胶子传递强核力，W 和 Z 玻色子传递弱核力，光子传递电磁力，而引力则由引力子( graviton) 传递，前三种力的传递子均早已被实验探测到[23]，引力波则直至2016年才由LIGO团队进行宏观星体的探测时发现[24][25][26][27]。

宏观星体间引力作用状态的变化会辐射引力波，微观世界的粒子间同样具有引力，其运动状态的变化因能量守恒亦将导致引力作用能以波子的形式辐射。微观粒子间引力作用强度仅为电磁作用的 $10^{-42}$～$10^{-36}$（质子/质子[23]、质子/电子、电子/电子间的比依次为 $10^{-36}$、$10^{-39}$、$10^{-42}$ 量级），所以引力对原子内电子电磁跃迁的影响可忽略不计。但电子的电磁跃迁却决定着引力能级状态的改变与引力波的辐射，只因引力与电磁作用力相比非常微弱，电子跃迁辐射的引力波子与物质间极弱的相互作用很难在实验中观察到。

以氢原子为例，其能级跃迁辐射光子的能量为：

$$\underbrace{2\pi\hbar\nu}_{n\to m} = E_n^e - E_m^e \tag{49}$$

式中$E_n^e$、$E_m^e$分别为氢原子第 n、m 级能级的电磁能，$2\pi\hbar\nu$ 为氢原子从第 n 能级跃迁到第 m 能级时所辐射光子的能量，电磁能$E_i^e$为

$$E_i^e = -\frac{1}{4\pi\varepsilon_0}\frac{e^2}{2r_i} \tag{50}$$

式中$r_i$为氢原子第 i 电磁能级的平均轨道半径，e 为电子电量。氢原子由电磁能级 n 跃迁到能级 m 时释放能量为$2\pi\hbar\nu$光子，同时引力作用能也由第 n 能级的$E_n^g$（第 n 能级的引力作用能记为$E_n^g$，后同）跃变为第 m 能级的$E_m^g$，根据能量守恒原理同样将辐射能量为$2\pi\varepsilon'\nu$的引力（波）子：

$$\underbrace{2\pi\varepsilon'\nu}_{n\to m} = E_n^g - E_m^g \tag{51}$$

式中$\varepsilon'$为质子（氢核）与电子间的引力作用量'常量'；定义$\varepsilon$为电子间引力作用量'常量'，因质子为电子质量的 1836 倍，所以$\varepsilon=1836^{-1}\varepsilon'$（本文暂未将中微子间引力作用计入讨论）。

因

$$E_i^g = -G\frac{m_p m_e}{2r_i} \tag{52}$$

$E_i^g$为氢原子处于第 i 电磁能级时的万有引力作用能，G 为万有引力常数，$m_p$ 为质子质量，$m_e$ 为电子质量。由(50)、(52)式得到同一轨道的电磁作用能与引力作用能之比为：

$$E_i^e\Big/E_i^g = \left\{\frac{1}{4\pi\varepsilon_0}\frac{e^2}{2r_i}\right\}\Big/\left\{G\frac{m_p m_e}{2r_i}\right\} \approx 10^{39} \tag{53}$$

电子由第 n 能级跃迁到第 m 能级时，电磁辐射子与引力辐射子能量之比：

$$\underbrace{2\pi\hbar\nu}_{n\to m}\Big/\underbrace{2\pi\varepsilon'\nu}_{n\to m} = \hbar/\varepsilon' = \left\{\frac{1}{4\pi\varepsilon_0}\frac{e^2}{2r_n} - \frac{1}{4\pi\varepsilon_0}\frac{e^2}{2r_m}\right\}\Big/\left\{G\frac{m_p m_e}{2r_n} - G\frac{m_p m_e}{2r_m}\right\} \approx 10^{39} \tag{54}$$

$$\Rightarrow \varepsilon' = 10^{-39}\hbar \tag{55}$$

将$\varepsilon=1836^{-1}\varepsilon'$、(55)式代入(34)式，得到引力作用图景下的不确定关系式

$$\Delta x \Delta p_x \geq \varepsilon/2 \approx 10^{-42}\,\hbar/2 \ll \hbar/2 \tag{56}$$

当$x$与$p_x$为高斯分布时，引力作用图景下微观粒子位置与动量的不确定度之积为：

$$\Delta x \Delta p_x = \varepsilon/2 \approx 10^{-42}\,\hbar/2 \ll \hbar/2 \tag{57}$$

在引力作用图景下，微观粒子共轭力学量不确定度关系式的下限仅为电磁作用图景中的$10^{-42}$，大大突破了关系式(1)的下限$\hbar/2$。引力作用图景描述下，力学量统计分布更集中，若以图像表示，分布曲线更锐利，因此引力作用图景可为描述微观粒子提供更为准确工具；同时也证明现行不确定性关系与量子理论仅适用于电磁作用图景下微观粒子力学状态的描述。

# 4 对不确定性原理及相关问题的简要探讨

根据上文的分析，以下就不确定性原理与物理实在及其相互关系进行简要讨论：

3.1 与 3.2 条的分析显示，现有量子理论是在电磁相作用图景下对微观粒子运动状态的统计描述，不确定性关系则是对这种统计描述特点的表征。海森堡提出不确定性原理时所作出的不确定性关系的极限与测量精度之间的兼容性判断，实际上与"单次测量可以预期的任何精度"无关，与微观粒子的位置/动量是否同时具有确定的值无关，也与微观物体的力学状态的客观性/确定性的存在性无关。基于不确定性原理而作出的微观粒子不能同时具有确定的位置与动量的断言，既不符合量子理论的出发点，也不符合微观粒子客观存在的物理现实。

宇观星体、宏观物体以及微观粒子，皆为外于人类的客观实在，都具有不依赖于意识、但可以被描述的物理状态。对于宇观星体、宏观物体，我们都可以依据现有理论对其运动状态作出完备的描述。那么，能否通过现有量子理论工具获得微观粒子力学状态的客观、确定性知识并进一步对其客观实在性与确定性做出判断？通过构造广义作用图景集合、根据量子理论作用量量子化思想、运用量子力学数学工具对微观粒子进行描述时，结果显示微观粒子波函数的形态在广义基本作用量'常量'趋向零极限的过程中同步演化。从波函数所具有的明确一致的演化终点，可获得微观粒子力学状态的真实、准确描述。因此，通过确定性极限判则可以明确断定：①波函数在演化起点（比如电磁作用图景）与演化过程（比如引力作用图景）中表现出来的、非统计诠释指称的量子几率性，实则是相应作用图景下微观粒子力学状态呈现相的统计概率；②不同作用图景下各波函数具有的共同一致的演化终点表明，微观粒子具有客观、确定、可描述的力学状态。

客观物理实在不依赖于人类意识，独立存在，那么如何理解作为客观物理实在的能量辐射子与具形的微观粒子所呈现的波粒二象性？3.3 的分析显示，对于微观粒子和场，没有相互作用（即$\varepsilon_x = 0$）就没有物理效应，没有物理效应就没有相应的物理呈现相（但$\varepsilon_x = 0$时可通过极限判则得到其客观物理属性的无扰描述），量子现象包括波粒二象性的呈现也同样如此。场在时域或空间的分布（有序的统计累积）呈现的波动性、能量吸收或辐射时表现的粒子（单元）性、粒子在时域与空间存在的个体奇异性-粒子性、统计描述（属粒子在作用图景中呈现的无时序累积）时呈现的波动性，以及场与粒子都存在的波粒二象性，皆是通过粒子与作用图景场之间的相互作用而得以呈现。没有粒子与场的相互作用，粒子不会显现波动性、场和波也不会显现粒子性；没有相互作用在时域或者空间存在的累积，就没有波动性的显现，粒子与波都不会有波粒二象性的呈现，波动性、粒子性以及波-粒二象性皆为互作用呈现相的特性。

物理世界是遵循因果决定性还是本质上的几率性与不确定？在量子力学形成之前，选择前者几乎是毫无异议；但在量子力学诞生尤其是不确定性原理提出之后，却成为困扰量子力学乃至物理学的一个根本性问题。由于 1930 年索尔维会议稳固确立了非统计诠释的主流正统地位，以及此后量子力学在诸多应用领域获得的巨大成功，哥本哈根学派否定因果决定性、失去对客观实在性的坚持所造成的影响已波及物理学之外的其它自然学科乃至于社会人文领域[30]。非统计诠释对实在性与因果决定性的放弃[1][30]受到爱因斯坦、劳厄、薛定谔、德布罗意以及后来的玻姆等少数几位坚持实在论观点物理学家强烈的质疑与反对[1][2]，尤其是面对玻尔直接提出"上帝不掷骰子"的爱因斯坦，终其一生都在对此进行拷问[1]。但由于长期以来非统计诠释一直占据着量子力学理论诠释的正统地位，致使绝大部分物理学家一度失去了对客观实在性、确定性与因果决定性的思考与坚持，以至于发生了时任国际力学联合会主席的 Sir James Lighthill 为之前以确定性误导公众而代表力学界同仁道歉[3][31]的一幕，更甚至于出现了在严肃的科学研究中居然相信意识决定实在、相信未来的测量行为会改变现在物理状态的现象[2]。究其根源，一方面由于缺乏在数学上对非统计诠释下的不确定性关系式进行深

度的检视分析、缺少对不同作用图景下微观粒子力学状态呈现相的考察比较、没有对量子理论取得巨大成就的应用领域及研究对象的统计属性与量子理论数学形式体系自身属性进行相关性考察，结果导致对非统计诠释的逻辑自洽性存在错误认识，对微观客体的实在性与非统计诠释理论断言的主观性之间、对量子客体力学状态的动态性/确定性与现行理论描述能力的有限性之间的关系存在模糊认识。另一方面，在方法论层面上，物理实在、物理理论与理论对物理实在的描述三者之间，若以理论为镜、则结果描述是像，镜、像所描述的对应体为物理实在。对于不确定性原理和量子力学的物理诠释，以玻尔海森堡为代表的哥本哈根学派只关注研究镜像、以像为实，爱因斯坦则始终坚持物理实在第一性。因此，就方法论而言，爱因斯坦与玻尔各自坚持物理诠释的范畴不同。不确定性原理与量子力学的物理诠释分歧之所以持续近百年，这也是根本原因之一。无可否认，量子理论仅是认识客观实在的工具，客观物理实在则既是理论度量描述的目标与对象、又是校验鉴定理论的终极标尺。

# 5 结论

本文的分析研究，揭示了不确定性原理非统计诠释的内在逻辑矛盾，明确了统计诠释的适用范围，解决了持续近百年的不确定性原理与量子力学物理诠释分歧问题。研究同时还揭示了微观粒子的力学状态具有客观性、确定性，发现了精确描述微观粒子确定力学状态的极限判则。

5.1 关于不确定性原理

（1）不确定性原理的非统计诠释存在逻辑矛盾、难以成立，非统计诠释下的量子力学理论不能逻辑自洽。

（2）统计诠释是不确定性原理与量子力学在电磁作用图景下唯一合乎逻辑的物理诠释。但若超出电磁作用图景，比如在引力作用图景中，现行不确定性原理的理论断言以及量子力学关于期望值之外的预言难以成立。

5.2 关于客观确定性

(1)极限判则显示微观粒子的力学状态具有客观确定性，"上帝不掷骰子"。

(2)客观确定性推论之一：现有波函数是电磁作用图景下微观粒子力学状态统计呈现相的希尔伯特相空间表述。对于不同的广义作用图景，相应波函数的数学形式在结构上一致——唯一差别在于以常数出现的量子化广义作用量'常量'，波函数可视为相应量子化基本作用量'常量'的因变量。

（3）客观确定性推论之二：量子力学的几率性并非微观粒子的内禀属性——非统计诠释指称的量子几率性，实质为微观粒子与作用图景场互作用呈现相的统计概率。

本文的研究结果，将有助于增进对微观世界量子物理现象的理解，亦可为考察量子信息技术的物理基础提供相应的理论参考。

# 6 致谢

# 参考文献

----------------------

# From the Uncertainty Principle to the Deterministic Rule
## --The falsification of the non-statistical interpretation of the uncertainty principle and the breakthrough of the statistical interpretation

Duan De-Long

(Aerospace Information Research Institute, Chinese Academy of Sciences; Beijing 100190, China)

E-mail: duandl@aircas.ac.cn; 13681112887@163.com

**Abstract:** It is well known that the quantum technology is developing rapidly, but the problem of divergence in the physical interpretation of quantum mechanics originating from the uncertainty principle has not yet been resolved. In order to clear the constraints and confusion of this situation for the further development of quantum technology, this article traces the original derivation and the physical meaning of the Heisenberg uncertainty principle, analyzes the Einstein photon−box thought-experiment, and studies the limits of the relationships under different action scenarios. Basing on the investigation and analysis, the two opposing physical interpretations of the uncertainty principle are studied, and the relationship between the lower limit of the relationship and whether the mechanical state of microscopic particles is objectively certain, the uncertainty principle and the scope of application of the statistical interpretation of quantum mechanics is studied under different action scenarios. By analyzing the statistical distribution of quantum mechanical quantities, the result of the destruction of the non-statistical interpretation uncertainty relation in the electromagnetic interaction scenarios is obtained; through analyzing of the photon box thought experiment, the logical contradiction of Bohr's argument was discovered; by examining the set of interaction scenarios, a description method for determining the mechanical state of microscopic particles was put forward; according to the analysis of the hydrogen atom transition radiation process, basing on the principle of conservation of energy, the result of the lower limit of the uncertainty relation in the gravitational scene is much smaller than that in the existing electromagnetic scene is obtained, which mean the lower limit of uncertainty principle is broken. The research in this article reveals the source of the individual probability of the quantum mechanics referred to by non-statistical interpretation and the inherent logical contradiction contained in the non-statistical interpretation itself, clearly gives the scope of application of statistical interpretation, and solves the divergent issues in the physical interpretation of quantum mechanics originating from the uncertainty principle. The research in this paper also reveals that the mechanical state of microscopic particles is objectively deterministic and discovers the method of limit criteria for determining its objective mechanical state, has received an affirmative answer to Einstein's "God does not play dice with the Universe". The research of this paper is helpful to enhance the understanding of quantum physical phenomena in the microscopic world, and we hope it could render some theoretical support to investigating the physical basis of quantum technology.

**Key words:** quantum mechanics; uncertainty principle (relation); Copenhagen interpretation of quantum mechanics; Photon-Box; graviton; quantum information

**PACS:** 03.65.Ca,03.65.Ta,01.70.+w, 01.65.+g, 95.85.Sz,03.67.-a

## 1. Introduction

Quantum mechanics(QM), as one of the two pillars of modern physics, has brought unprecedented success in its application along the way of science. It has not only reshaped physics in a sense, but also has caused substantial change in human society.[1][2][3][4] Nevertheless, the disputes over the physical interpretation of QM originated from the uncertainty principle(UP) have existed for a long time.[2][5] And simultaneously, with the rapid deployment of quantum information technology, the problem of divergence in the physical interpretation of QM focusing on UP has severely affected the development of quantum technology.

The uncertainty relationship(UR), also known as UP, is called by Bohr and Pauli one cornerstone of QM. The proposal of the UP is honored to be a major achievement in science. The essential difference between QM and classical mechanics lies in the fundamental difference between the probability and indeterminacy of QM and the determinism of classical mechanics, and caused the decades long debate between Einstein and Bohr. From the logical consistency of the UP, which is equivalent to the logical self-consistency of QM, to the completeness of QM targeted by the EPR theory, they debated for many years. Until today, the divergence between the probabilistic description of the microcosm in the non-statistical interpretation (NSI) of QM and the decisive and precise description of classical causality insisted by Einstein still remains.[1][2][3][4][5][6][7][8]

The physics community has no objection to the mathematical derivation of the uncertainty relation. The fundamental difference lies in the physical interpretation of the uncertainty principle. The physical interpretation is divided into the following two schools: NSI and Statistical interpretation(SI). [1]

**NSI/Bohr-Heisenberg interpretation.**

The UR is a description of individual quantum system. The canonical conjugate variables of a quantum system cannot be accurately determined at the same time, and the lower limit of the product of uncertainty is $\hbar/2$.[6] This is the mainstream interpretation of QM, also known as the orthodox interpretation of QM or the Copenhagen interpretation. Its essence is to adhere to the individual probability of QM and to abandon the causal determinism that physics has adhered to since the Galileo era.

**Statistical Interpretation(SI)**

The UR describes the ensemble of identically prepared quantum systems, and the lower limit of the product of the standard deviations of the statistical distribution of the regular conjugate variables is $\hbar/2$. The essence of this interpretation insisted by Einstein and Schrödinger is to insist on and not to give up the view of classic causal decisive and precise description.

After the Solvay-controversy between Bohr and Einstein, some physicists such as Heisenberg, Schrödinger, De Broglie, Magenau, Blokhintsev, Zhang Hanliang, Guan Hong, Huang Xiangyou, and Ozawa M., have made useful explorations in this area, but still failed to eliminate the fundamental differences in its physical interpretation and the problem of divergence of theoretical interpretation is still not resolved. In view of the troubles caused by this situation on the layout and development of quantum technologies, we reexamine the original derivation and physical meaning of the uncertainty relation, conduct probability analysis on the NSI, and look into its mathematical and physical logical self-consistency. We also review and analyze Bohr's argument about the photon-box, and explore the logical consistency of the argument. And then by analyzing the evolution characteristics of the uncertainty relation and the mathematical form of the wave function as the virtual basic action 'constant' of the generalized action scene tends to the zero limit, we investigate the source of quantum probability, and analyze the possibility of accurately describing the mechanical state of microscopic particles. Finally, we deduce the corresponding mathematical relationship in the gravitational scene trying to ascertain the application range of the current QM and the UR.

## 2. The Proposal of UP and NSI

The construction of the mathematical formal system of QM is ahead of its physical interpretation. In order to adress the two issues of the QM mathematical formal system: ① Does quantum theory allow the position and velocity of a particle to be determined with the limited precision at a given moment? ② Whether the accuracy allowed by the theory is compatible with the best accuracy obtained in the experimental measurement, Heisenberg published "On the Intuitive Content of Quantum Theory's Kinematics and Mechanics" in March 1927[1][6], proposed UP and through the mathematical formal system of QM derived the relational formula (1):

$$\Delta q \Delta p \geq \hbar/2 \tag{1}$$

where $\Delta q \equiv \sqrt{\overline{(\Delta q)^2}} = \sqrt{\overline{(x-\bar{x})^2}}$、$\Delta p \equiv \sqrt{\overline{(\Delta p)^2}} = \sqrt{\overline{(p_x - \overline{p_x})^2}}$, through thought experiments such as γ-ray microscope and electron single-slit position determination in [6], the uncertainty relation is physically interpreted.[1][2] Defining the uncertainty relation is the limitation of quantum theory on the accuracy of simultaneous measurement of conjugate mechanical quantities, constitutes the core of NSI. NSI emphasizes that "any exact observation result (accuracy, precision) must comply with the UR" [8], the physical source of adherence to the accuracy limit is the inevitable interference caused by the measurement operation[6][9][10][11][12][13], and it is the UP that protects the logical consistency of QM[10].

NSI is currently the standard interpretation acceped by most Chinese and foreign QM/theoretical mechanics textbooks, natural philosophy monographs and popular science books on QM theory[1][2][6][14][15][16], and it is also the theoretical basis of quantum communication and quantum computing (In the following, the relational expression $\Delta q \Delta p \geq \hbar/2$ (2) is used to express the relational expression of the UP under NSI).

## 3. Analysis and discussion on the UP

In order to solve the fundamental divergence in physical interpretation of UP and study its scope of applicationin in this chapter, we first analyze the consistency between the statistical distribution of the quantum mechanical quantities of microscopic particles and the UR under NSI, and examine the logical consistency of the UR (section 3.1); we then analyze Bohr's argument for the photon-box thought experiment, and study the logical consistency of Bohr's argument (Section 3.2); we construct a set of action scenes and examine the description of the motion state of microscopic particles when the basic interaction 'constant' tends to the limit (Section 3.3.1/2), and finally we analyze and discuss the expression of the UR in the gravitational action scene (Section 3.3.3).

### 3.1 Mathematical analysis of NSI of uncertainty relation

In this section we will analyze the coincidence between the statistical distribution of quantum mechanical quantities and the uncertainty relation under NSI, and examine the logical consistency of the NSI of the uncertainty relation.

### 3.1.1 Analysis of the coincidence between the statistical distribution of quantum mechanical quantities and NSI

In order to analyze the conformity between the statistical distribution of quantum mechanical quantities and the uncertainty relation of NSI, the definition is the same to hypothesis of [6], let the functions $\psi(q')$ and $\Psi(p')$ be the probability amplitudes of the distribution of the mechanical quantities of electrons q and p respectively. Suppose q is the Gaussian distribution, and its probability amplitude is $\psi(q')$, then p is also a Gaussian distribution, and its probability amplitude is $\Psi(p')$ from the Fourier transform. The expression of the probability amplitudes is as follows:

$$\psi(q') = \frac{1}{(2\pi)^{1/4}\sqrt{(\Delta q)}} \cdot e^{-\frac{(q'-\bar{q})^2}{4(\Delta q)^2} - \frac{2\pi i}{h}\bar{p}q'}$$

$$\Psi(p') = \int \frac{1}{\sqrt{h}} e^{\frac{i}{\hbar}q'p'} \psi(q') dq' \qquad (3)$$

① Define $\Delta q' = q' - \bar{q}$ as the deviation of any single measurement q from $\bar{q}$, which represents the single measurement accuracy of $q$. Let $q'$ satisfy $q' \in (\bar{q} - \Delta q, \bar{q} + \Delta q)$, that is, the probability of $(q', p')$ for $\Delta q' < \Delta q$ is $\eta_{q'}$

$$\eta_{q'}=\{\int_{\bar{q}-\Delta q}^{\bar{q}+\Delta q}|\psi(q')|^2 dq'\}/\{\int |\psi(q')|^2 dq'\} \tag{4}$$

Where Δq is the Δq defined when q is Gaussian distribution in formula (2).

From the property of definite integrals, if the function $f(x) \geq 0$ holds on [a, b], then the following formula

$$\int_a^b f(x)\,dx \geq 0$$

must be established, and because $|\psi(q')|^2 > 0$ while $q' \in (\bar{q}-\Delta q, \bar{q}+\Delta q)$ and $\int|\psi(q')|^2 dq' = 1$, then get:

$$\eta_{q'}=\{\int_{\bar{q}-\Delta q}^{\bar{q}+\Delta q}|\psi(q')|^2 dq'\}/\{\int |\psi(q')|^2 dq'\} > 0 \tag{5}$$

② Since $q$ and $p$ are Gaussian distributions, $\Delta q$ and $\Delta p$ in formula (2) satisfy $\Delta q \Delta p = \hbar/2$, Where Δp is the Δp defined when $p$ is Gaussian distribution in formula (2). Let the probability of $(q', p')$ satisfying $\{p' \in (\bar{p}-\Delta p, \bar{p}+\Delta p), \Delta q \Delta p = \hbar/2\}$ be $\eta_{p'}$, the probability of $(q', p')$ satisfing $\{q' \in (\bar{q}-\Delta q, \bar{q}+\Delta q), p' \in (\bar{p}-\Delta p, \bar{p}+\Delta p), \Delta q \Delta p = \hbar/2\}$ is $\eta_{q'p'}$, then:

$$\eta_{p'} = \int_{\bar{p}-\Delta p}^{\bar{p}+\Delta p} \Psi(p')\Psi^*(p') dp'$$
$$= \int_{\bar{p}-\Delta p}^{\bar{p}+\Delta p} \left(\int \frac{1}{\sqrt{h}} e^{\frac{i}{\hbar}q'p'} \psi(q')dq'\right)\left(\int \frac{1}{\sqrt{h}} e^{\frac{i}{\hbar}q'p'} \psi(q')dq'\right)^* dp' > 0 \tag{6}$$

$$\eta_{q'p'} = \int_{\bar{p}-\Delta p}^{\bar{p}+\Delta p} \left(\int_{\bar{q}-\Delta q}^{\bar{q}+\Delta q} \frac{1}{\sqrt{h}} e^{\frac{i}{\hbar}q'p'} \psi(q')dq'\right)\left(\int_{\bar{q}-\Delta q}^{\bar{q}+\Delta q} \frac{1}{\sqrt{h}} e^{\frac{i}{\hbar}q'p'} \psi(q')dq'\right)^* dp'$$
$$= \int_{\bar{p}-\Delta p}^{\bar{p}+\Delta p} \left|\int_{\bar{q}-\Delta q}^{\bar{q}+\Delta q} \frac{1}{\sqrt{h}} e^{\frac{i}{\hbar}q'p'} \psi(q')dq'\right|^2 dp' \tag{7}$$

From the formula (7) and the property of definite integral, we can get the following formulas

$$\eta_{q'p'} = \int_{\bar{p}-\Delta p}^{\bar{p}+\Delta p} \left|\int_{\bar{q}-\Delta q}^{\bar{q}+\Delta q} \frac{1}{\sqrt{h}} e^{\frac{i}{\hbar}q'p'} \psi(q')dq'\right|^2 dp' > 0 \tag{8}$$

$$\eta_{q'p'} > 0 \tag{9}$$

③ From formulas (7) and (8), we can get the following formula (9) while the condition is met $\{q' \in (\bar{q}-\Delta q, \bar{q}+\Delta q), p' \in (\bar{p}-\Delta p, \bar{p}+\Delta p), \Delta q \Delta p = \hbar/2\}$

$$\Delta q' \Delta p' < \hbar/2 \tag{10}$$

Then formula (9 holds

$$\eta_{q'p'} > 0 \tag{9}$$

The result of probabilistic analysis contradicts the predictions of NSI. The NSI of "the UR that any exact observation (precision) must comply with"[9] formula (2) is invalid and untrue.

### 3.1.2 Discussion on the Failure of NSI of UP

The reason why the probabilistic analysis results of 3.1.1 contradict the NSI predictions lies in the following:

When [6] deriving the UR, Heisenberg used not only the statistical wave function of the coordinates and momentum but also the transformation of the representation and the SI of the transformation function. His qualitative analysis of the γ-ray microscope also used implicit statistical relationships-Einstein-de.Broglie relationship. If there is no statistical wave function and de.Broglie relationship, the UR cannot be obtained. As a direct inference of the mathematical formal system of QM, that is, a conclusion obtained through the Dirac-Jordan transformation theory, the uncertainty relation (1) represents the constraint relationship between the standard deviation of the conjugate mechanics quantity , and it is the equivalent mathematical expression of SI[1][2][3][6].

Bohr believes that the clarity of the pure mathematical form itself has little value for a physical proposition, and a complete physical explanation should be absolutely higher than its mathematical formal system[1][14]. After Heisenberg derives formula (1), the mean square error $\sqrt{(\Delta q)^2}$ and $\sqrt{(\Delta p)^2}$ in the formula is defined as the uncertainty of the electronic position and momentum to express the measurement accuracy separately, and the mathematical relationship is interpreted as the constraint relationship (2) of the accuracy of measuring the conjugate mechanical quantity at the same time.[6]

The mean square error $\sqrt{(\Delta q)^2}$, $\sqrt{(\Delta p)^2}$ in the UR (1) represents the dispersion degree of the conjugate mechanical quantities $q$ and $p$, and presents the characteristics of statistical distribution of $q$ and $p$ values[1][2][3], both of which must be obtained through statistical analysis of a large number of high-precision measurement results, and cannot be obtained only by a single measurement.[5] But Heisenberg clearly uses it to express the measurement accuracy in the relation (2)—the accuracy is the deviation of the obtained value of $q$ or $p$ from the true (expected) value, which belongs to the individual characteristics of the single measurement. Heisenberg reinterpretes Δq and Δp in the relation (1) as measurement accuracy, which causes the loss of the inherent mathematical meaning of standard deviation in statistical theory, and consequently, the mathematical and physical meanings of Δq and Δp have changed from root mean square error to uncertainty at last to measurement accuracy , and caused the connotations  inconsistent. Undoubtedly, this violates the principle of logical consistency. Therefore, in the probability analysis of 3.1.1, the result that the relationship (2) is destroyed and the NSI is not valid has been obtained

At the same time, NSI has not received any direct verification support from real physical experiments so far.[1][5] Not only that, it is worthy to notice that the reanalysis (See details [17] and article 3.2 in this paper) of several thought experiments such as Heisenberg γ-ray microscope and Einstein photon box that provided physical interpretation support for it, also get the research result that the uncertainty relation (2) is destroyed, which is consistent with the mathematical proof in 3.1.1.

As the equivalent mathematical expression of the SI of QM, the UR has no restriction on the accuracy of a single measurement that is different from the source of the standard deviation, nor can any limitation be made on the product $\Delta q'\Delta p'$ of the single measurement accuracy of the mechanical quantities $p$ and $q$. There are no upper and lower limits on the accuracy of a single measurement. SI does not deny that microscopic particles can have a definite position and momentum at the same time. Moreover, the UR in the SI sense is not unique to QM, and there is no lack of corresponding existence in classical theory.[3][17][18][19][20]

NSI lacks logical consistency, neither the UR nor the current QM theory can properly describe the mechanical state of microscopic particles under NSI. The UR and the current QM theory can only properly describe the mechanical state of microscopic particles under SI.

### 3.2 Re-analyzing the photon box thought experiment

Bohr's demonstration of the photon box thought experiment at the Sixth Solvay Conference successfully refuted Einstein's view that NSI cannot be logically self-consistent, and established the mainstream and orthodox position of NSI in QM[1][15]. This section conducts a retrospective analysis of Bohr's argument and examines its logical consistency.

### 3.2.1 A review of the debate between Bohr and Einstein on the photon-box

In order to refute Bohr's NSI of the UP, Einstein proposed a photon box thought experiment at the Solvay Conference in Brussels in October 1930. He believed that the two measurements of photon emission time T and photon energy E were independent of each other and did not interfere with each other, the measurement accuracy did not restrict each other, therefore the experiment will destroy the time-energy UR, thus proving that the UP and the NSI of QM are not logically self-consistent. But Bohr immediately disproves it as follows: [1][8][16]

When the photon with energy E escapes, the photon box obtains an upward momentum (impulse) $p$ from the photon emission.

$$p \leq T \frac{E}{c^2} g \tag{11}$$

Where T is the time required for the photon-box to reach a new equilibrium position from the beginning of the photon release. The uncertainty $\Delta p$ of the momentum $p$ obtained by the photon-box is

$$\Delta p \leq T \frac{\Delta E}{c^2} g \tag{12}$$

Before and after the photon escapes, the difference $\Delta x$ between the two equilibrium positions of the photon box is related to $\Delta p$, and $\Delta x$ and $\Delta p$ satisfies the formula (13)

$$\Delta p \geq \frac{\hbar}{\Delta x} \tag{13}$$

From (12) and (13), we can get:

$$\hbar \leq T \frac{\Delta E}{c^2} \Delta x g \tag{14}$$

The uncertainty of T , $\Delta T$ is determined by the position uncertainty $\Delta x$(the speed of the clock changes due to changes in gravitational potential), according to the gravitational red-shift equation (15):

$$\frac{\Delta T}{T} = g \frac{\Delta x}{c^2} \tag{15}$$

Substituting (15) into (14), Bohr gives the time-energy UR

$$\Delta E \cdot \Delta T \geq \hbar \tag{16}$$

Einstein could not refute Bohr's argument, Bohr's conclution was accepted by the attendees [1][8][16] and it is considered as "a particularly rigorous analysis of the complexity of QM." [1][15]. Since then, Einstein did not publicly question the logical self-consistency of the UP and QM basing on NSI, but in fact he never agreed with NSI. Moreover, from the end of the Solvay Conference in 1930 to the end of his life in 1962, Bohr himself has repeatedly returned to thinking about the photon-box argument[1][8].

### 3.2.2 Reanalysis of Bohr's Argument

Bohr's argument at that time was considered to be "very clear description of the physical basis for the correct interpretation of QM". So is it logically flawless and can truly prove the logical self-consistency of NSI? The following analyzes from Bohr's argument logic and the logical inference     the conclusion of Bohr's argument.

### 3.2.2.1 Bohr's argument has the following problems

① There are errors in the derivation from (11) to (12)

The following is defined by the momentum (impulse)

$$p = T \frac{E}{c^2} g \tag{17}$$

Ignoring high-order small quantities, the following formula can be obtained:

$$\Delta p = \Delta T \frac{E}{c^2} g + T \frac{\Delta E}{c^2} g \tag{18}$$

Therefore, Bohr's demonstration to derive momentum uncertainty $\Delta p$ expression (12) from momentum p expression (11) is not valid.

② The inconsistency of the meaning of mathematical symbols in the process of Bohr's argument

In the " Before and after … satisfies the formula (13) ", $\Delta x$ represents the difference between the two equilibrium positions of the photon-box; the $\Delta x$ in "the uncertainty of T... equation (15)" is used to express the position uncertainty. In fact, the difference between the two equilibrium positions of the photon-box and its position uncertainty are not the same.

③ The logical loop problem of Bohr's argument

In order to prove the time-energy UR, Bohr promises that the position-momentum UR    is established in "Before and after … satisfies the formula (13)", and introduced its variant Δp≥h/Δx. This is self-citation in the argumentation logic and belongs to a circular argument.

④ The question of whether the "Before and after … satisfies the formula (10)" is established

Supposing the photon energy is E, the mass-energy equation can be used to obtain the difference Δx between the two equilibrium positions before and after the photon is released by the photon-box as follows

$$\Delta x = \frac{E}{kC^2} g \quad (19)$$

The position difference Δx depends on the photon energy E and the elastic coefficient k of the traction spring of the photon-box. The position difference Δx has nothing to do with the unknown source Δp. Both Δx and Δp are not restricted by equation (13), therefore, the "Before and after … satisfies the formula (13)" on which Bohr's argument is based is not valid.

⑤ The problem of uncertainty of the transition time between the two equilibrium positions while weighing photon-box

The gravitational red-shift causes the clock to slow down, record $\Delta T_b$ as the change of the weighing time $T_b$ ($T_b$ is the time T used in Bohr's argument in Section 3.2.1 at formula (11) ~ (16), and the required time the photon box opens and closes the shutter to release photons is recorded as $T_a$) of the photon-box due to the red-shift effect, which can be known from the gravitational red shift formula:

$$\Delta T_b = T_b \frac{g\Delta x}{C^2} \quad (20)$$

$\Delta T_b$ is the time change that can be accurately calculated by the above formula, It is not a time to be uncertainty.

Recording the position uncertainty of the photon box as $\Delta x'$, which is determined by the limiting accuracy of the equilibrium position of the photon box measured when weighing, Obviously $\Delta x' < \Delta x$ holds for a position measurement with physical meaning. The position uncertainty $\Delta x'$ will cause the uncertainty of the gravitational potential change, which in turn will lead to the uncertainty of the speed of the clock and watch, and the uncertainty of the time $T_b$ measurement should be

$$\Delta T_b' = T_b \frac{g\Delta x'}{C^2} \quad (21)$$

Because of $\Delta x' < \Delta x$, the uncertainty of time $T_b$ measurement is $\Delta T_b'$, so $\Delta T_b' < \Delta T_b$. Therefore, the time uncertainty on which Bohr's argument is based is incorrect.

### 3.2.2.2 A counter-argument to Bohr's argument according to Bohr's argumentation logic

① Assuming that Bohr's proof of the time-energy UR of the photon box thought experiment is established, E and $T_b$ are Gaussian distributions, the following formula is established:

$$\Delta E \cdot \Delta T_b = \hbar \quad (22)$$

② Let $\Delta x_1$ be the displacement within the time $T_a$ of the photon released by the photon-box, suppose the photon box opens the shutter at time $t_1$, the photon box closes the shutter at time $t_2$, and the time when the photon box reaches a new equilibrium position and finally completes the weighing is $t_3$, $t_1<t_2<t_3$, obviously $\Delta x_1$ {$\Delta x_1 = \int_{t_1}^{t_2} v(t)dt$, $v(t) > 0$} is less than the total displacement of the photon box during weighing Δx {$\Delta x = \int_{t_1}^{t_2} v(t)dt + \int_{t_2}^{t_3} v(t)dt$}, that is, $\Delta x_1 < \Delta x$.

From the gravitational redshift formula (15), the time $T_a$ ($T_a = t_2 - t_1$) of the photon released by the photon box can be obtained, and the time change caused by the gravitational redshift recording as $\Delta T_a$

$$\Delta T_a = g \frac{\Delta x_1}{c^2} T_a \quad (23)$$

The uncertainty of time Ta measurement caused by the position uncertainty of the photon box is $\Delta T_a'$:

$$\Delta T_a^{'} = g\frac{\Delta x^{'}}{C^2}T_a \qquad (24)$$

As the photon-box release time $T_a$ is extremely short, the time $T_b$ ($T_b=t_3-t_1$) of the photon-box weighing process can be arbitrarily long, that is, $T_a < T_b\{\Delta x_1 < \Delta x\}$; so $\Delta T_a^{'} < \Delta T_b^{'}\{\Delta T_a < \Delta T_b\}$ and the following formula is obtained:

$$\Delta E \cdot \Delta T_a^{'} < \Delta E \cdot \Delta T_b^{'} < \Delta E \cdot \Delta T_b \qquad (25)$$

$$\{\Delta E \cdot \Delta T_a < \Delta E \cdot \Delta T_b \qquad (26)\}$$

Since item ① has assumed that E and $T_b$ satisfies the Gaussian distribution, that is, $\Delta E \cdot \Delta T_b = \hbar$ holds, it is predicted that the time-energy UR of the photon released by the photon-box to the distant place satisfies the following formula

$$\Delta E \cdot \Delta T_a^{'} < \hbar \qquad (27)$$

$$\{\Delta E \cdot \Delta T_a < \hbar \qquad (28)\}$$

The result of (27) { (28) } destroys the time-energy UR (16) and (22), so Bohr's argument is self-contradictory (Note: This Article {... } is the deduction of Bohr's argumentation logic based on [15]).

Bohr's demonstration of the photon box thought experiment at the Solvay Conference in 1930, was once considered to be a successful model of proving the logic self-consistent of the NSI of the UP and QM[1][15]. However, the above logical analysis shows that Bohr's argument at the time is obviously not valid, and the NSI of the UP and QM cannot be logically self-consistent (Note: the study to the literature [1] and [8] also got the same results).

**3.3 Discussion of the non-electromagnetic interaction scene**

Article 3.1&3.2 proves that NSI is not logically self-consistent, and SI is a reasonable physical interpretation of the UP and QM. Then, under the SI, can the current UP and QM be able to universally describe the mechanical state of microscopic particles in various interaction scenarios? For this reason, this section will first derive the standard deviation constraint relation of the conjugate mechanical quantity under the generalized action scenario; then, by constructing the action scenario set, examine the description of the microscopic particle motion state when the basic action "constant" takes the limit; finally, according to the principle of conservation of energy, by introducing the "constant" of the gravitational interaction, in the analysis of the electromagnetic radiation process of the hydrogen atom, the investigation of the expression of the UR and the survey of the possibility of breaking the limit $\hbar/2$ in the gravitational scenario will be realized.

**3.3.1 The Derivation of the Restricted Relations of the Standard Deviation of Conjugate Mechanics under the Generalized Action**

In order to realize the description of the mechanical state of microscopic particles under the generalized interaction scenario, based on Born's commutative relation quantization thought and the basic mechanical quantity Heisenberg equation, refer to [6][21][22], set the undetermined interaction scene as $X(\varepsilon_x)$, and define its basic action 'constant' as $\varepsilon_x$. In the context of $X(\varepsilon_x)$, any particle is in a finite space, and the three coordinate axes of the space are x, y, z. In order to simplify the analysis, only one-dimensional situation is considered, thus only the *x* coordinate of the three spatial coordinates is discussed. Making the particle move along the *x* direction, $p_x$ is the momentum of the particle. Assuming the coordinate wave function is $\psi(x)$, the momentum wave function is $\Psi(p_x)$, and the standard deviation is used to express the uncertainty of coordinates and momentum

$$\Delta x = \sqrt{\overline{(x-\bar{x})^2}}, \quad \Delta p_x = \sqrt{\overline{(p_x-\overline{p_x})^2}}$$

Assuming that the average values of *x* and $p_x$ are both zero, we consider the following inequality that is obviously true

$$\int_{-\infty}^{\infty} \left|\alpha x\psi + \frac{d\psi}{dx}\right|^2 dx \geq 0 \tag{29}$$

where α is an arbitrary real constant, due to

$$\int x^2 |\psi|^2 dx = (\Delta x)^2 \tag{30}$$

$$\int \left(x \frac{d\psi^*}{dx}\psi + x\psi^* \frac{d\psi}{dx}\right) dx = \int x \frac{d|\psi|^2}{dx} dx = -\int |\psi|^2 dx = -1 \tag{31}$$

$$\int \frac{d\psi^*}{dx}\frac{d\psi}{dx} dx = -\int \psi^* \frac{d^2\psi}{dx^2} dx = \frac{1}{\varepsilon_x^2} \int \psi^* \hat{p}_x^2 \psi dx = \frac{1}{\varepsilon_x^2}(\Delta p_x)^2 \tag{32}$$

and so

$$\alpha^2 (\Delta x)^2 - \alpha + \frac{1}{\varepsilon_x^2}(\Delta p_x)^2 \geq 0 \tag{33}$$

The parabolic quadratic trinomial (33) of α is greater than or equal to zero for all α, so its discriminant must be a non-positive real number, so

$$\Delta x \Delta p_x \geq \frac{1}{2}\varepsilon_x \tag{34}$$

The smallest possible value of the product of $\Delta x \Delta p_x$ is $\frac{1}{2}\varepsilon_x$, and the wave function takes the following form:

$$\psi = \frac{1}{(2\pi)^{1/4}\sqrt{(\Delta x)}} exp\left(\frac{i}{\varepsilon_x} p_0 x - \frac{x^2}{4(\Delta x)^2}\right) \tag{35}$$

where $p_0$ and $\Delta x$ are constants, $\Delta x$ is the standard deviation, we gain the probability function of Gaussian distribution as follows

$$|\psi|^2 = \frac{1}{\sqrt{(2\pi)} \cdot \Delta x} exp\left(-\frac{(x-\bar{x})^2}{2(\Delta x)^2}\right) \tag{36}$$

The wave function in the momentum representation is

$$\Psi(p_x) = \frac{1}{\sqrt{2\pi\varepsilon_x}} \int_{-\infty}^{\infty} \psi(x) e^{-(i/\varepsilon_x)xp_x} dx \tag{37}$$

the probability is

$$|\Psi|^2 = \frac{1}{\sqrt{(2\pi)} \cdot \Delta p_x} \times exp\left(-\frac{(p_x - \bar{p_x})^2}{2(\Delta p_x)^2}\right) \tag{38}$$

Momentum probability distribution $|\Psi|^2$ is also Gaussian, and its standard deviation is $\Delta p_x = \varepsilon_x/2\Delta x$, that is, the standard deviation constraint relation of $x$ and $p_x$ is:

$$\Delta x \Delta p_x = \frac{1}{2}\varepsilon_x \tag{39}$$

**3.3.2 Analyzing the determinism of the microscopic particle mechanics state**

In order to investigate whether a deterministic description of the mechanical state of microscopic particles can be obtained, we construct the generalized action scene set {Xi} (i=1,2,3...n...), the element sequence of the basic interaction 'constant' set {εxi} corresponding to {Xi} has the relationship of the following formula

$$\varepsilon_{x1} > \varepsilon_{x2} > \varepsilon_{x3} > \ldots > \varepsilon_{xn} > \ldots \to 0 \tag{40}$$

which is

$$\lim_{n \to \infty} \varepsilon_{xn} = 0 \tag{41}$$

Assuming that the position $x_i$ of the microscopic particle under the action scene $X_i$ is Gaussian distribution, then the momentum $p_{x_i}$ is also Gaussian distribution and the product of the standard deviations of $x_i$ and $p_{xi}$ satisfies the following formula

$$\Delta x_i \Delta p_{xi} = \frac{1}{2}\varepsilon_{xi} \tag{42}$$

According to equations (40) and (42), the following formulas holds

$$\Delta x_1 \Delta p_{x1} > \Delta x_2 \Delta p_{x2} > \Delta x_3 \Delta p_{x3} > \ldots \Delta x_n \Delta p_{xn} > \ldots \to 0 \tag{43}$$

which is:

$$\lim_{n \to \infty} \Delta x_n \Delta p_{xn} = 0 \tag{44}$$

The following equations can be obtained by the definition of action[28]_

$$\lim_{n \to \infty} \Delta x_n = 0 \tag{45}$$

$$\lim_{n \to \infty} \Delta p_{xn} = 0 \tag{46}$$

When n→∞, the momentum and position wave functions (24) and (25) of microscopic particles evolve into two δ functions: [28][29]

$$\lim_{\Delta x_n \to 0} |\psi| = \lim_{\Delta x_n \to 0} \left| \frac{1}{(2\pi)^{1/4}\sqrt{(\Delta x_n)}} exp\left(-\frac{i}{\varepsilon_{xn}}\overline{p_{xn}}x_n - \frac{x_n^2}{4(\Delta x_n)^2}\right) \right| = \delta(x = \bar{x}, \Delta x_n = 0) \tag{47}$$

$$\lim_{\varepsilon_{xn} \to 0} |\Psi| = \lim_{\varepsilon_{xn} \to 0} \left| \frac{1}{(2\pi)^{1/4}\sqrt{\varepsilon_{xn}}} \int_{-\infty}^{\infty} \psi(x_n) exp\left(-\frac{i}{\varepsilon_{xn}} x_n p_{xn}\right) dx_n \right| = \delta(p_x = \overline{p_x}, \varepsilon_{xn} = 0) \tag{48}$$

The two formulas (47) and (48) that degenerate into the δ function in the generalized action scene description analysis, the position and momentum values determined by the microscopic particles can be obtained, and the mechanical state of the microscopic particles can be obtained accurately. From the limit criterion adopted above, it can be concluded that microscopic particles have an objective, definite, and accurately described mechanical state.

**3.3.3 Expression of Uncertainty Relations in the Context of Gravitation**

The SI is a reasonable physical interpretation of the UP and QM. Can it properly describe the mechanical state of microscopic particles in the context of gravity? According to the principle of conservation of energy, this section analyzes the luminescence process of hydrogen atoms by introducing the 'constant' of the gravitational interaction, and realizes the expression of the UR of the gravitational action scene, and examines the applicability of existing QM description and the Heisenberg UR formula (1).

The four basic forces that exist in nature are all transmitted by particles: gluons transmit strong nuclear forces, W and Z bosons transmit weak nuclear forces, photons transmit electromagnetic forces, and gravitational forces are transmitted by gravitons. The first three transmitters of these forces have been detected by experiments, [23] and gravitational waves were not discovered until 2016 when the LIGO team was conducting macroscopic star detection. [24] [25] [26] [27]

The changes in the state of gravitational action between macroscopic stars will radiate gravitational waves, and particles in the microscopic world also have gravitational forces, and the changes in their motion states will also cause gravitational energy to radiate in the form of waves due to energy conservation. The intensity of gravitational action is only $10^{-42} \sim 10^{-36}$ of the electromagnetic action (the ratio of proton/proton[23], proton/electron, electron/electron is on the order of $10^{-36}$, $10^{-39}$, $10^{-42}$), therefore, the influence of gravity on the electromagnetic transition of electrons in the atom is negligible. However, the electromagnetic transition of electrons determines the change of gravitational energy level and the radiation of gravitational waves. It is only because gravity is so weak compared with electromagnetic force that the extremely weak interaction between gravitons (gravitational waves of electron transition radiation) and matter is too difficult to be observed by the current experimental instruments.

Taking hydrogen atom as an example, the energy of photon radiated by its energy level transition is

$$\underbrace{2\pi\hbar\nu}_{n \to m} = E_n^e - E_m^e \tag{49}$$

Where $E_n^e$ and $E_m^e$ are the electromagnetic energy of the Nth and Mth energy levels of the hydrogen atom respectively, $2\pi\hbar\nu$ is the energy of the photon radiated by the hydrogen atom from the Nth energy level to the Mth energy level. The electromagnetic energy $E_i^e$ is

$$E_i^e = -\frac{1}{4\pi\varepsilon_0} \frac{e^2}{2r_i} \tag{50}$$

Where $r_i$ is the average orbital radius of the Ith electromagnetic energy level of the hydrogen atom, and e is the electron charge. While the hydrogen atom transitions from the electromagnetic energy level n to the energy level m, and the energy released is $2\pi\hbar\nu$ photons, and the gravitational energy also changes from $E_n^g$ at the Nth energy level (the gravitational action energy of the Nth energy level is denoted as $E_n^g$, the same later ) to $E_m^g$ at the Mth energy level, according to the principle of conservation of energy, will also radiate gravitons (gravitational wave) with energy $2\pi\varepsilon'\nu$:

$$\underbrace{2\pi\varepsilon'\nu}_{n \to m} = E_n^g - E_m^g \quad (51)$$

Where $\varepsilon'$ is the pseudo-elementary action 'constant' of the gravitational interaction between the proton (hydrogen nuclei) and the electron (define $\varepsilon$ as the elementary action 'constant' of the gravitational action between electrons. Because the proton is 1836 times the mass of the electron, so $\varepsilon=1836^{-1}\varepsilon'$, this article does not include the gravitational interaction between neutrinos in the discussion)

Since

$$E_i^g = -G\frac{m_p\, m_e}{2r_i} \quad (52)$$

$E_i^g$ is the gravitational energy when the hydrogen atom is at the Ith electromagnetic energy level, G is the gravitational constant, $m_p$ is the mass of the proton, and $m_e$ is the mass of the electron. From equations (50) and (52), the ratio of electromagnetic energy to gravitational energy in the same orbit is

$$E_i^e / E_i^g = \{\frac{1}{4\pi\varepsilon_0}\frac{e^2}{2r_i}\} / \{G\frac{m_p\, m_e}{2r_i}\} \approx 10^{39} \quad (53)$$

The ratio of the energy of electromagnetic radiation to gravitational radiation when the electron jumps from the Nth energy level to the Mth energy level is

$$\underbrace{2\pi\hbar\nu}_{n \to m} / \underbrace{2\pi\varepsilon'\nu}_{n \to m} = \hbar/\varepsilon' = \{\frac{1}{4\pi\varepsilon_0}\frac{e^2}{2r_n} - \frac{1}{4\pi\varepsilon_0}\frac{e^2}{2r_m}\} / \{G\frac{m_p\, m_e}{2r_n} - G\frac{m_p\, m_e}{2r_m}\} \approx 10^{39} \quad (54)$$

$$\Rightarrow \varepsilon' = 10^{-39}\hbar \quad (55)$$

Substituting $\varepsilon=1836^{-1}\varepsilon'$ and Eq.(55) into Eq.(34), then obtain the UR formula in the context of gravitational action

$$\Delta x \Delta p_x \geq \varepsilon/2 \approx 10^{-42}\, \hbar/2 \ll \hbar/2 \quad (56)$$

When $x$ and $p_x$ are Gaussian distributions, the product of the uncertainty of the microscopic particle position and momentum is

$$\Delta x \Delta p_x = \varepsilon/2 \approx 10^{-42}\, \hbar/2 \ll \hbar/2 \quad (57)$$

Under the gravitational action scenario, the lower limit of the UR of the conjugate mechanics of microscopic particles is much more lower than that of the electromagnetic action scenario—the former is only $10^{-42}$ of the latter, which greatly breaks the restriction of relation (1); if these are represented by an image, the statistical distribution of mechanical quantities is more concentrated and the distribution curve is sharper than that of the electromagnetic action scenario. Therefore, the context of gravitational interaction can provide a more accurate tool for describing microscopic particles. It also proves that the current UR and quantum theory are only applicable to the description of the mechanical state of microscopic particles in the context of electromagnetic interaction.

## 4. A brief discussion on the uncertainty principle and physical reality

Based on the analysis completed above, the following further discusses the UP and physical reality and their relationship:

The analysis of Articles 3.1 and 3.2 shows that the current quantum theory is a statistical description of the motion state of microscopic particles in the context of electromagnetic interaction, and the UR is a representation of the characteristics of this statistical description. The compatibility judgment between the limit of the UR and the measurement accuracy made by Heisenberg when he proposed the UP has nothing to do with the assertion of "any accuracy that can be expected from a single measurement". This limited relationship has nothing to do with whether the position/momentum of the microscopic particle has a definite value at the same time, and it has nothing to do with the objectivity/certainty existence of the mechanical state of the microscopic object. The assertion made on the basis of the UP that microscopic particles cannot objectively have a definite position and momentum at the same time does not conform to the starting point of quantum theory or the physical reality of the objective existence of microscopic particles.

Universe stars, macroscopic objects, and microscopic particles are all objective realities outside of human beings. They all have physical states that are not dependent on consciousness and can be described at the same time. For cosmoscopic stars and macroscopic objects, we can make a complete description of their motion state based on existing theories. Then, is it possible to obtain objective and deterministic knowledge of the mechanical state of microscopic particles through the existing quantum theory tools and make further judgments on its objective reality and certainty? By constructing a set of action scenarios, according to the quantum theory of action quantization, while using the mathematical tools of QM to describe microscopic particles, it is discovered that the form of the microscopic particle's wave function evolves synchronously as the elementary action 'constant' approaches the zero limit, the definite evolution end point of the wave function can obtain a true and accurate description of the mechanical state of microscopic particles; therefore, through this deterministic limit criterion, the results received clarify that the non-statistically interpretative quantum probability of the wave function in the evolution starting point and evolution process is actually the statistical probability of the microscopic particle mechanics state in the corresponding action scene; and the wave function has a consistent evolution end point under different action scenarios, indicating that microscopic particles have an objective, definite, and describeable mechanical state.

The objective physical reality does not depend on human consciousness and exists independently. So how to understand the wave-particle duality presented by the energy radiators and the concrete microscopic particles as the objective physical reality? The analysis of section 3.3 shows, for physical reality such as microscopic particles and energy radiators, there is no physical effect without interaction(while $\varepsilon_x = 0$ ), and there is no corresponding physical appearance without physical effect (But when $\varepsilon_x$ =0, an undisturbed description of its objective physical properties can be obtained through the deterministic limit criterion). The same is true for quantum phenomena of energy radiators including wave-particle duality. The volatility of the field distribution in the time domain or space (ordered statistical accumulation), the particle characteristic of energy absorption or radiation, the individual singularity of the particle in the time domain and space called particle nature , statistical description (the non-chronological accumulation of the particles in the action scene), as well as the wave-particle duality that exists in both energy radiators and particles, are all presented through the interaction between the particles and the action scene field. Without the accumulation of interactions in the time domain or space, there will be no volatility, and there will be no wave-particle duality between particles and waves. Wave, particle, and wave-particle duality are all characteristics of interaction.

Does the physical world obey causal determinism, or is it inherently probabilistic and uncertain? Before the formation of QM, there was almost no objection to choosing the former; while after the exploitation of QM, especially after the UP was proposed, it became a fundamental problem that plagued QM and even the science of physics itself. Due to the solid establishment of the orthodox position of NSI in the theory of QM after the Solvay Conference in 1930, as well as the great success of QM in a comprehensive applications, the impact of the denial of causality and the loss of adherence to objective reality has spread to other natural disciplines besides physics and even to social and humanistic fields. The NSI of the abandonment of realism and causality[1][30] was strongly questioned and opposed by a few physicists who insisted on realism, such as Einstein who directly proposed that "God does not play dice with the universe" for opposing Bohr, had been interrogating it untill to the end of his life.[1] However, because NSI has always occupied the orthodox position of theoretical interpretation of QM for a long period of time in history, most physicists once lost their thinking and insistence on the objective reality, certainty, and causality. As a result, Sir James Lighthill, the chairman of IUTAM at the time, apologized to the public on behalf of colleagues in the mechanics circle for misleading the public with certainty,[3][31] and even in serious scientific research there is a phenomenon which believes that consciousness determines reality actually and that future measurement behaviors will change the current physical state. [2] Why is there such a situation? On the one hand, as the previous analysis shows, the root cause lies in the lack of mathematical in-depth examination and analysis of the UR under NSI, a lack of comparisons and studies of the mechanical states of microscopic particles under different action scenarios, and no investigation of the correlation between the statistical properties of quantum theory's application fields and the properties of the quantum theory mathematical formal system itself. As a result, there is a misunderstanding of the logical self-consistency of NSI, a vague understanding of the relationship between the reality of microscopic objects and the subjectivity of NSI theoretical assertions, and fuzzy knowledge of the relationship between the dynamics/determinism of the mechanical state of quantum objects and the finiteness of the current theoretical description ability. On the other hand, as far as methodology is concerned, between physical reality, physical theory and the description of physical reality by theory, theory can be regarded as a mirror, results described as images, and mirrors and images correspond to physical reality. For the physical interpretation of the UP and QM, the Copenhagen School represented by Bohr and Heisenberg only focused on the study of mirror images and took the image as reality, while Einstein always insisted on the first nature of physical reality. Therefore, at the methodological level, the two interpretations have different categories. This is also one of the fundamental reasons why the divergence of physical interpretation of the UP and quantum theory lasted for nearly a hundred years. It is undeniable that quantum theory is only a tool to measure and describe objective reality, while objective physical reality is not only the target and object of theoretical understanding of measurement and description, but also a ruler for verifying and identifying theories.

## 5. Conclusions

The research in this article revealed the inherent logical contradiction of the NSI of the UP, clarified the scope of application of SI, and resolved the divergence of the physical interpretation of the UP and QM that have lasted for nearly a hundred years. The research also revealed that the mechanical state of microscopic particles is objective and deterministic, and found the limit criterion that could accurately describe the mechanical state of microscopic particles.

**5.1 About the uncertainty principle**

(1) The NSI of the UP has logical contradictions and is difficult to establish. The QM theory under the NSI is also not logically self-consistent.

(2) SI is the only logical physical interpretation of UP and QM in the context of electromagnetic interaction, if it exceeds the context of electromagnetic interaction, such as in the gravitational action scene, the assertion of the existing uncertainty relation and the prediction of QM theory are not valid except for the expected value.

**5.2 About objective certainty**

(1) The analysis in this article shows that the mechanical state of microscopic particles is objectively certain，"God does not play dice with the Universe."

(2) One of the objective deterministic inferences: The existing wave function is a statistical representation of the mechanical state of microscopic particles in the context of electromagnetic interaction, expressed in the Hilbert phase space. For different action scenarios, the mathematical structure of the corresponding wave function is the same—the only difference is the quantized elementary action 'constant' that appears as a constant, and the wave function is the dependent variable of the quantized elementary action 'constant' of the corresponding action scenario.

(3) The second corollary of objective certainty: the probability of QM is not the intrinsic property of microscopic particles. The quantum probability of NSI refers to the statistical probability of the interaction between microscopic particles and the interaction scene.

The research results of this paper is helpful to enhance the understanding of quantum physical phenomena in the microscopic world, and could render some corresponding theoretical support to investigating the physical basis of quantum technology.

# Acknowledgements

The author would like to thank Professor Xin Hao of Nanjing University of Posts and Telecommunications, Associate Professor Li Xue-Wen of Beijing Institute of Technology, Researcher Shan Huan-Yan, Researcher Li Shi, Researcher Tan Rong-Qing, Researcher Wang Yong, Researcher Yin Sheng-Yi from the CAS Aerospace Information Research Institute, and Mr. Li Yi-Yong from the USA for their assistance to this paper; and thanks to Professor  Zhang Cai-Bo of Shangdong University(Weihai), CHEN Ling-Xiao Senior Engineer of Beijing University of Posts and Telecommunications for the work in text modification and entering.

--------------------